# Multiple modes of a single spin torque oscillator under the non-linear region

Satoshi Sugimoto,[1,*] Shuichi Iwakiri,[2] Yusuke Kozuka,[1] Yukiko Takahashi,[1] Yasuhiro Niimi,[2,3] and Kensuke Kobayashi,[2,5] and Shinya Kasai,[1,4]

[1]*Research Center for Magnetic and Spintronic Materials, National Institute for Materials Science (NIMS), 1-2-1 Sengen, Tsukuba 305-0047, Japan*

[2]*Department of Physics, Graduate School of Science, Osaka University, Osaka 560-0043, Japan*

[3]*Center for Spintronics Research Network (CSRN), Graduate School of Engineering Science, Osaka University, Osaka 560-8531, Japan*

[4]*JST, PRESTO, 4-1-8 Honcho, Kawaguchi, Saitama 332-0012, Japan*

[5]*Institute for Physics of Intelligence (IPI) and Department of Physics, Graduate School of Science, The University of Tokyo, Tokyo 113-0033, Japan*

**Abstract**

**A numerical investigation is conducted for a single spin-torque oscillator under the non-linear region. A large angle precession triggers the generation of multiple modes without any feedbacked circuits and/or magnetic couplings with neighboring oscillators. Our simulations show that a single eigenmode of a given spin-torque oscillator can trigger up to six discrete modes as the sideband modes. These findings will offer the new functionality to the STO for developing the spintronic logic circuits.**

*Corresponding author: SUGIMOTO.Satoshi@nims.go.jp

Spin-torque oscillator (STO) [1] is an auto-oscillator on the nanometer scale driven by the spin toque under the charge current. It has attracted considerable attention with prospects of a wide range of applications including the frequency signal generator [2], signal modulation [3], spin-wave generation [4], and neuromorphic computing [5].

STO is normally governed by a single Kittel mode in a linear auto-oscillatory region, while the non-linear STO driven by the phase-locking [6] enables us to engineer the frequency modulation. The sideband modulation [7,8,9,10,11,12] is a typical analog frequency modulator where equally spaced multiple modes are excited around the fundamental mode. The schemes of the sideband modulation are similar to multi-mode laser systems in terms of a delay-line oscillator. The delay in the latter case is due to the optical cavity while the delay is brought by the external cavity in the case of feedback STO, e.g., the external radio-frequency sources [9,11]. In addition, a recent experimental work report the reflection from the transmission line can also trigger sideband-type spectrum through the filtering effects [12]. However, these external cavities mostly require huge circuit components, including the phase shifters and mixtures, centimeter-sized cables for signal reflection. These device extensions prevent downsizing of the whole oscillatory system and abandon the great potential in nanoscale communication system of STO.

In this paper, we numerically show a different scheme to induce the multiple modes only by using a single radio-frequency source without any special device structures. We have conducted micromagnetic simulations to investigate the non-linear self-sustained dynamics for a single STO setup based on the spin Hall effect [12,13,14,15,16,17,18]. The large angle precession under weak perpendicular magnetic anisotropy results in the periodical perturbations in the oscillation trajectories. This leads to the multiple modes via the non-linear frequency and amplitude modulation (NFAM) [8]. Importantly, this type phenomenon is triggered only by intrinsic magnetization dynamics of a single STO without requiring any external sources to assist the synchronization. Further downsizing is possible down to the size limit of the single STO nanostructure, possibly several tens of nanometers [19].

A schematic image of the present model system is shown in Fig. 1. We assume that the single STO is driven by the spin Hall effect asserted throughout a heavy metal electrode with strong spin-orbit interaction (Fig. 1). We focus on the magnetization dynamics of the spatially averaged free layers with typical material parameters of CoFeB [20] for simplicity. The micromagnetic simulations are carried out by using the open-source micromagnetic code, the finite difference package MuMax3 [21]. The motions of the spins are calculated by solving Landau-Lifshitz-Gilbert equation [22,23] with spin-transfer and spin Hall torque contributions;

$$\frac{\partial \mathbf{m}}{\partial t} = -\gamma(\mathbf{m} \times \mathbf{H}_{\text{eff}}) + \alpha \left( \mathbf{m} \times \frac{\partial \mathbf{m}}{\partial t} \right) + \mathbf{T}_{\text{I}} + \mathbf{T}_{\text{SH}}, \quad (1)$$

where $\mathbf{m}$ denotes the normalized free-layer magnetization and $M_s \mathbf{m} = \mathbf{M}$ with the saturation magnetization $M_s (= 1.41 \times 10^4 \text{ Oe})$. $\gamma (= 1.76 \times 10^{11} \text{ Hz/T})$ is the gyromagnetic ratio, and $\alpha (= 0.012)$ is the Gilbert damping constant. $\mathbf{H}_{\text{eff}}$ is the effective field including the external magnetic field $H$, the magnetic anisotropy field $\mathbf{H}_{\text{an}} = \frac{2K_u}{M_s}(\mathbf{u} \cdot \mathbf{m})\mathbf{u}$, with the first order uniaxial anisotropy constant $K_u$ and a unit vector indicating the anisotropy direction $\mathbf{u}$, and the Heisenberg exchange field $\mathbf{H}_{\text{ex}} = 2\mu_0 \frac{A}{M_s} \Delta \mathbf{m}$, with the exchange stiffness constant $A(= 2.84 \times 10^{-11} \text{ J/m})$. $\mathbf{T}_{\text{I}}$ counts the spin-polarized current contribution given by $\mathbf{T}_{\text{I}} = (\gamma \hbar P I_{\text{MTJ}}/2eM_s V)\mathbf{m} \times (\mathbf{m} \times \mathbf{p})$, and $T_{SH}$ is the spin Hall torque given by $\mathbf{T}_{SH} = (\gamma\hbar\theta_{SH}J_{SH}/2eM_sd)[\mathbf{m}\times(\mathbf{m}\times\hat{\mathbf{y}}) - \beta\mathbf{m}\times\hat{\mathbf{y}}]$, where $\mathbf{p}$ is the spin polarizer vector, $\hbar$ is the Dirac constant, $e$ is the elementary charge, $P(=0.57)$ is the spin polarization of CoFeB [20], $I_{MTJ}(=50\ \mu A)$ is the perpendicular DC current passing through STO, $V$, and $d$ are the volume and thickness of the free-layer, $\theta_{SH}(=-0.12)$ is the spin Hall angle of Ta [24], and $J_{SH} = I/(w_{HM}d_{HM})$ is the current density in the heavy metal electrode with $w_{HM}$ and $d_{HM}$ in its width and thickness. The complex behavior of the field-like spin Hall torque coefficient $\beta$ [25] will be crucial for investigating the magnetization switching process [26], but should not be critical for discussion about excitation modes. Here $\beta$ is set to be $\beta = 1$, for simplicity [27].

The STOs in the present calculation have an elliptical shape with the major and minor axes of 160 nm and 80 nm, respectively, with the free-layer thickness of $d = 1.3$ nm, allocated on the heavy metal electrode with $(w_{HM}, d_{HM}) = (1\ \mu m, 7\ nm)$. These values are based on the experimental work by Iwakiri et al. [12]. The discretization cell size is set at $2 \times 2 \times 1.3\ nm^3$. The in-plane external magnetic field $H = 200$ Oe is applied along $\phi = 60°$ direction, and $\mathbf{p}$ is fixed to be parallel to it.

Figure 2(a) shows the time evolution of the $z$ component magnetization $m_z$ in a linear oscillation region (top panel: $I = 0.91$ mA) and in a non-linear oscillation region (bottom panel: $I = 1.06$ mA). The magnetization continuously oscillates with a single frequency in the linear oscillation region, which is typical of STO. In the non-linear oscillation region, on the other hand, the precession angle gradually increases as a function of time. This transient state continues about 120 ns, and then magnetization dynamics results in a non-linear steady state in which a steady oscillation with a slow beating of precession angle starts to appear.

The top and bottom panels of Fig. 2(b) show the spectral densities obtained by fast Fourier transform (FFT) of $m_z$ in the linear and the non-linear oscillation regions, respectively. While the spectral density in the linear region shows a single peak at 695 MHz, the spectral density in the non-linear exhibits multiple peaks at 115 MHz, 490 MHz, 600 MHz, and 715 MHz. In addition, two tiny sideband peaks are observed at 230 MHz 375 MHz as discussed in detail below.

The time evolutions of the first and second fundamental modes (600 MHz and 115 MHz) are plotted in the top and middle panels of Fig. 3(b), accompanied by the full non-linear oscillation track in the bottom panel. As guided by the gray areas, the beating frequency coincides with the second fundamental mode 115 MHz. Also, frequencies of other four peaks in the spectra (230 MHz, 375 MHz, 490 MHz, and 715 MHz) are constructed by making difference and/or doubling of these two fundamental frequencies, i.e., $230\text{ MHz} = 115\text{ MHz} \times 2$, $375\text{ MHz} \cong 600\text{ MHz} - 115\text{ MHz} \times 2$, $490\text{ MHz} \cong 600\text{ MHz} - 115\text{ MHz}$, and $715\text{ MHz} = 600\text{ MHz} + 115\text{ MHz}$. These sideband modes can be provided from the general solutions of the non-linear universal auto-oscillator model [28], and have a strong analogy with the difference frequency generation and the second harmonic generation in non-linear optics [29].

Let us introduce the analytical approach developed by Consolo et al. [30] to describe the multiple mode generation. The output signal of STO can be denoted as $c(t) = A_c \cos(2\pi f_c t)$ with the oscillation amplitude $A_c$ and the base frequency $f_c = 600\text{ MHz}$ in the universal model. We can take into the phenomenological beating frequency $f_m = 115\text{ MHz}$ by writing $A_c \rightarrow A_c \cos(2\pi f_m t + \varphi)$ having the phase difference $\varphi$. Then the entire output wave can be written in the form of $c(t) = A_c(t)\cos[\theta(t)]$ using the time dependent amplitude ($A_c(t)$) and phase ($\theta(t)$) components, which can be generally categorized as the combined NFAM signal [8]. The peak positions of such NFAM spectrum can be obtained as $f = f_c^l \pm l f_m$, where $f_c^l$ is the central frequency and $l$ is the positive integer labeling sideband order. This NFAM scenario can reproduce all non-linear modes of the spectrum in Fig. 3 (a) by using the central frequency as $f_c^l = 600$ MHz or 115 MHz ; $f = 600 \pm 115 l_1$ MHz (375 MHz, 490 MHz, 715 MHz) and $f = 115 \pm 115 l_2$ MHz (230 MHz) with $l_1, l_2 = 0,1,2$. In these manners, these multiple modes are regarded to be equivalent up to the 2nd order of sideband modes of NFAM.

We next discuss the origin of the two different fundamental modes under the non-linear processes. The three-dimensional trajectories of the linear ($I = 0.91$ mA) and non-linear oscillations ($I = 1.06$ mA) are plotted in Fig. 4(a). Owning to the weak perpendicular anisotropy $K_u = 7.55 \times 10^5$ J/m$^3$, the magnetization dynamics is no longer free from the finite shape-induced in-plane anisotropy of the thin film ($K_i$, a red vector in Fig. 4(a)). The precession axis is then strongly tilted from the perpendicular axis ($K_p$, a black vector) [31,32]. An application of the in-plane external field $H$ (a blue vector) also tilts the azimuthal angle of the precession axis from the $x$-axis. As a consequence, the magnetization oscillates around the intermediate position where its azimuthal and elevation angles of the precession axis are located at $(\theta, \phi) \sim (30°, 30°)$.

Once the large-angle precessions are excited at such an intermediate position, the non-linear trajectories will be affected by the complicated triaxial-type anisotropy energy profile induce by $(K_i, K_p, H)$. This leads to the fluctuation of the precession angle as plotted by the red lines in Fig. 4(a), in which the non-linear oscillation shows different trajectories for every period of the fundamental mode at 600 MHz.

Figure 4(b) shows time evolutions of the precession axis of the free-layer magnetization at the linear and non-linear oscillations obtained by averaging the trajectories for one period of the fundamental modes. While the precession axis settles on the identical position during the linear oscillation, it draws the elliptical orbits in the non-linear oscillation with $\sim 8.70$ ns ( $= 115$ MHz ) in its average periodicity. This fluctuation of the precession axis triggers the second fundamental mode. It should be noted that the present results indicate that such multiple modes are purely inherent in the magnetization dynamics of a common cylinder-shaped STO nanopillar, and are free from any external circuits demonstrated so far [9,11].

The conditions of the multiple modes are addressed by calculating spectral densities as functions of electric current $I$, external field $H$, and perpendicular anisotropy $K_u$ in Fig. 5(a)~(c). The electric current $I$ dependence of spectral density is presented in Fig. 5 (a). The spectra show a linear single-mode $700 \pm 5$ MHz under $I \leq 0.94$ mA followed by the transient region with the nonlinear red shift (the red dotted lines), and then enter into the non-linear multiple modes above $I \geq 1.00$ mA. The widths and amplitudes of the peaks of the multiple modes are barely affected by $I$ under $I < 1.20$ mA. Interestingly, this result has a strong analogy with the recent experimental observation of the sideband-type spectrum [12], which implies the intrinsic relevance of the NFAM. Once the current exceeds $I > 1.20$ mA, the magnetization starts continuous switching processes.

The external in-plane field $H$ dependence under $I = 1.06$ mA is introduced in Fig. 5(b). The clear locking behavior is observed under $160 \text{ Oe} \leq H < 220 \text{ Oe}$. The first fundamental mode (~600 MHz) shows a small dependence on $H$ while the second fundamental mode (~115 MHz) is barely affected by $H$. These locking behaviors are easily corrupted under too small or too large field $H$ since they are attributed to a competition between the perpendicular and in-plane anisotropy energies. As results, single peak spectra are observed at $H < 160$ Oe and $H > 220$ Oe.

The perpendicular anisotropy $K_\text{u}$ dependence under $I = 1.06$ mA is shown in Fig. 5(c). Among the three modulation parameters $(I, H, K_\text{u})$, the locking condition for $K_\text{u}$ is confirmed to lies especially in the narrow region; $7.52 \times 10^5 \text{ J/m}^3 \leq K_\text{u} \leq 7.58 \times 10^5 \text{ J/m}^3$, around 0.5% deviation of $K_\text{u}$ would break the locking. Under the stronger $K_\text{u}(> 7.58 \times 10^5 \text{ J/m}^3)$, the precession axis has been barely affected by in-plane anisotropy $K_\text{i}$ and applied field $H$, hence the free-layer magnetization would precess in the way of the usual perpendicular STO. While under the weaker $K_\text{u}(< 7.50 \times 10^5 \text{ J/m}^3)$, STO property would be close to that of the in-plane oscillator, and the frequency of the fundamental peak decrease down to 600 MHz range. In other words, fine-tuning of $K_\text{u}$ may help to find out these NFAM conditions experimentally, such as the voltage-controlled magnetic anisotropy [33].

Finally, we introduce these non-linear effects on the output power of STO. The current dependence of the output power $P_\text{o}$ can be written as [9]

$$P_\text{o} = \left[\frac{R_\text{T}}{(R_\text{T}+R_\text{MTJ})^2}\right](0.5\Delta RI)^2 \text{var}(m_x), \quad (2)$$

where $\text{var}(m_x)$ is the variance of average value of $m_x$, $R_\text{T}$, $R_\text{MTJ}$, and $\Delta R$ denote the termination resistance, magnetic tunneling junction resistance, and magnetoresistance between parallel and antiparallel status, respectively. We set reference values as $R_\text{T} =$ 50 Ω, $R_{\text{MTJ}} = 3850$ Ω, and $\Delta R = 2100$ Ω after the experimental report about the non-linear multiple mode nucleation in single STO [12]. Figure 5(d) shows the current $I$ dependence of the output power $P$. The transient region from the linear precession to the non-linear precession is guided by the red dotted lines same as Fig. 5(a). The output power $P$ shows gradual increase in the linear region ($I < 0.94$ mA), and notable increase is observed once the NFAM is conducted. In NFAM region ($I > 1.00$ mA), $P$ shows a further increase and eventually reaches almost ten times enhancement compared to the linear output power. These results indicate the intrinsic sideband modulation possess the potential advantage not only to the downsizing, but also to improve an active device power following the typical phase-locking phenomena.

In summary, we have conducted numerical studies to induce the multiple modes of STO inherent in its magnetization dynamics. By using free-layer with the weak perpendicular magnetic anisotropy, NFAM by the beating mode leads to the difference frequency generation and the second harmonic generation, which are accompanied with a sharp increase of the output power. Since these non-linear processes require neither extrinsic radio frequency sources nor complicated device structure, they would be potentially usable at any design of the circuit. These results may open a new route for developing a new class of non-linear STO systems.

## ACKNOWLEDGEMENTS

This work was partially supported by Japan Society for the Promotion of Science (JSPS) KAKENHI grant nos. JP17K18892, JP18J20527, JP19H05826, JP19H00656,

JP16H05964, and JP26103002, and JST, PRESTO grant no. JPMJPR18L3, Japan. The authors acknowledge M. Takahagi for technical help.

## DATA AVAILABILITY

The data that support the findings of this study are available within this article.

**Figure Captions**

FIG. 1. Schematic image of a configuration of spin Hall STO with in-plane fixed layer and free-layer. A black vector denotes the spin polarizer layer (i.e. the fixed layer), and a red vector denotes the free-layer. The normalized magnetization vector **m** of free-layer magnetization is illustrated in the right inset in the polar coordinate representation with the azimuthal and elevation angles $(\phi, \theta)$.

FIG. 2. (a) The time evolution of the $z$ components of the free-layer magnetization for the linear oscillation (top panel: $I = 0.91$ mA) and the non-linear oscillation (bottom panel: $I = 1.06$ mA). The non-linear steady oscillation process is guided by gray colored area. (b) The spectral density of $m_z$ for the linear oscillation (top panel: $I = 0.91$ mA) and the non-linear oscillation (bottom panel: $I = 1.06$ mA). The spectral densities in the non-linear regime are deduced from results in the steady oscillation region (120ns $\leq t \leq$ 240 ns).

FIG. 3. (a) The spectral density of $m_z$ for the non-linear oscillation process obtained at 300 ns $\leq t \leq$ 500 ns. Peaks at 115 MHz (the second fundamental mode: solid blue line), 230 MHz (dotted blue line), 375 MHz (dotted green line), 490 MHz (solid green line), 600 MHz (the first fundamental mode: solid black line), and 715 MHz (solid green line) are guided, respectively. (b) The time evolutions of the first fundamental mode (top panel), second fundamental mode (middle panel), and $z$ components of the free-layer magnetization (bottom panel). Results of the former two modes are calculated independently from simulations, to visualize the two major frequencies in the bottom panel.

FIG. 4. (a) The real-space trajectories of the linear oscillation ($I = 0.91$ mA, black line) and the non-linear oscillation ($I = 1.06$ mA, red line). Time evolutions from $300$ ns $\leq t \leq 320$ ns are plotted. The directions of perpendicular anisotropy $K_\mathrm{p}$, in-plane anisotropy $K_\mathrm{i}$, and external field $H$ are plotted by black, red, and blue arrows respectively. (b) The real space trajectories of the precession axis at the linear oscillation (black symbol) and the non-linear oscillation (red symbol). Each point of trajectories is obtained by integrating free-layer magnetizations for the periods of the fundamental frequencies; 695 MHz for the linear oscillation and 600 MHz for the non-linear oscillation. The current amplitude and windows of time evolutions are identical with those of Fig. 4(a).

FIG. 5. The spectral density of $m_z$ as functions of (a) the electric current $I$, (b) external field $H$, and (c) perpendicular anisotropy $K_\mathrm{u}$. The red dotted lines in (a) indicates the transient region to the non-linear frequency and amplitude modulation (NFAM) condition. The white dotted lines in (b) and (c) guide the NFAM conditions. (d) The current dependence of the output power $P$ obtained by Eq. (2). The gray region indicates the transient region to the NFAM condition.

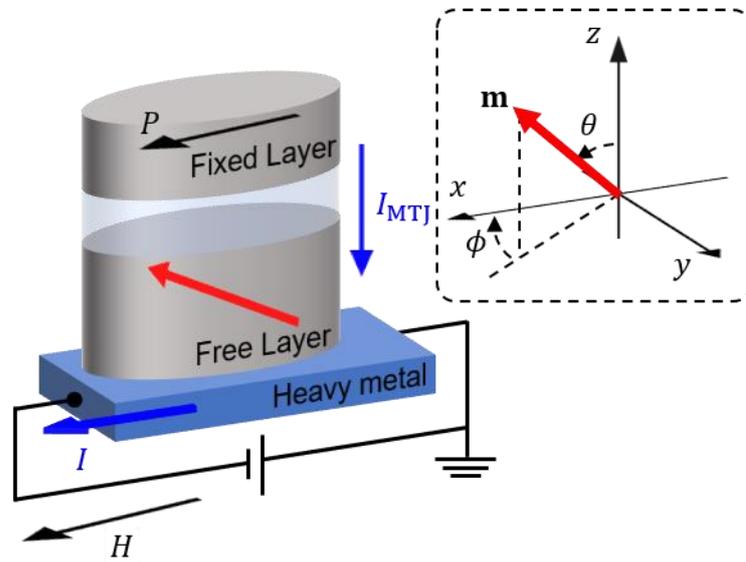

Fig. 1 Sugimoto *et al*.,

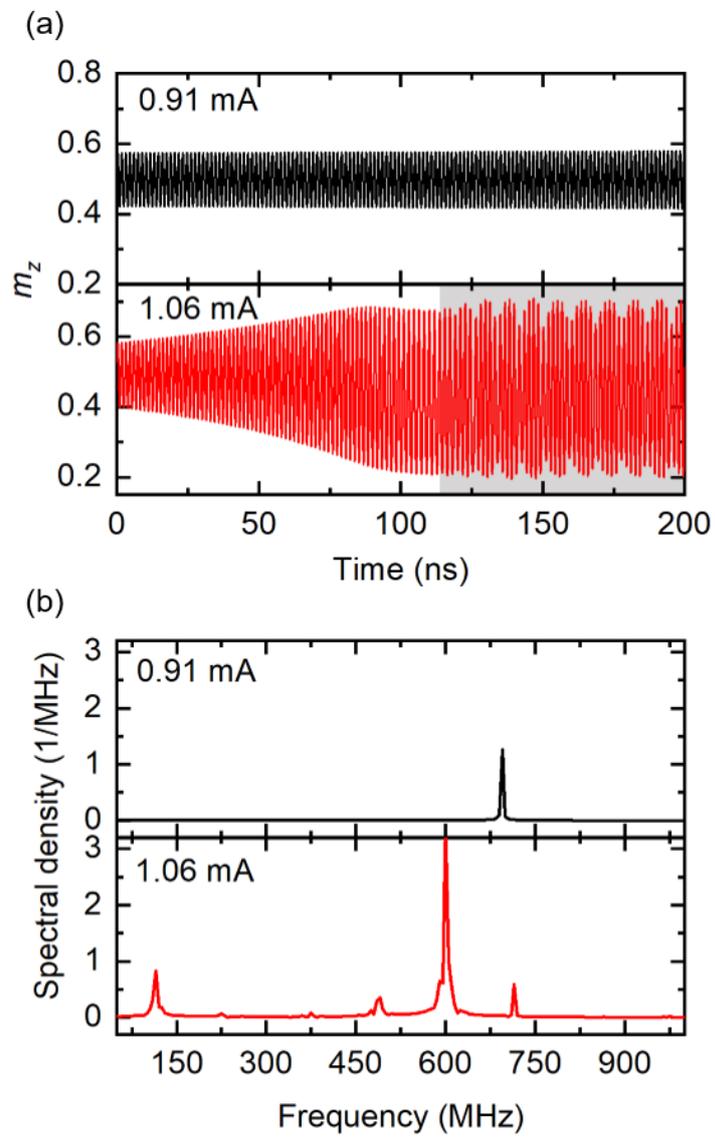

Fig. 2 Sugimoto *et al*.,

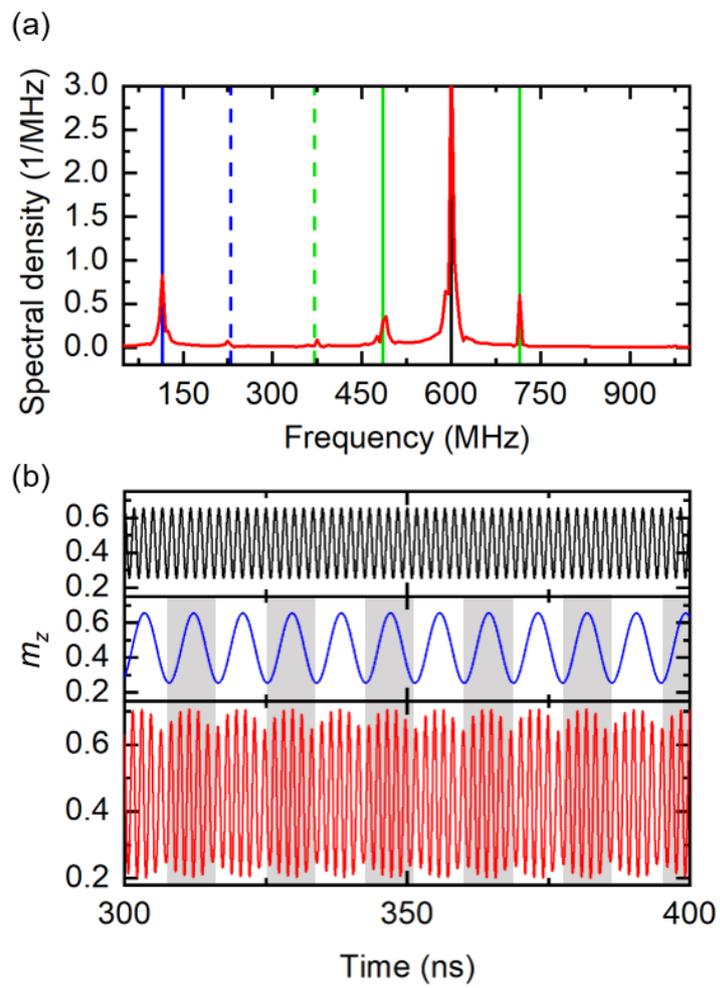

Fig. 3 Sugimoto *et al*.,

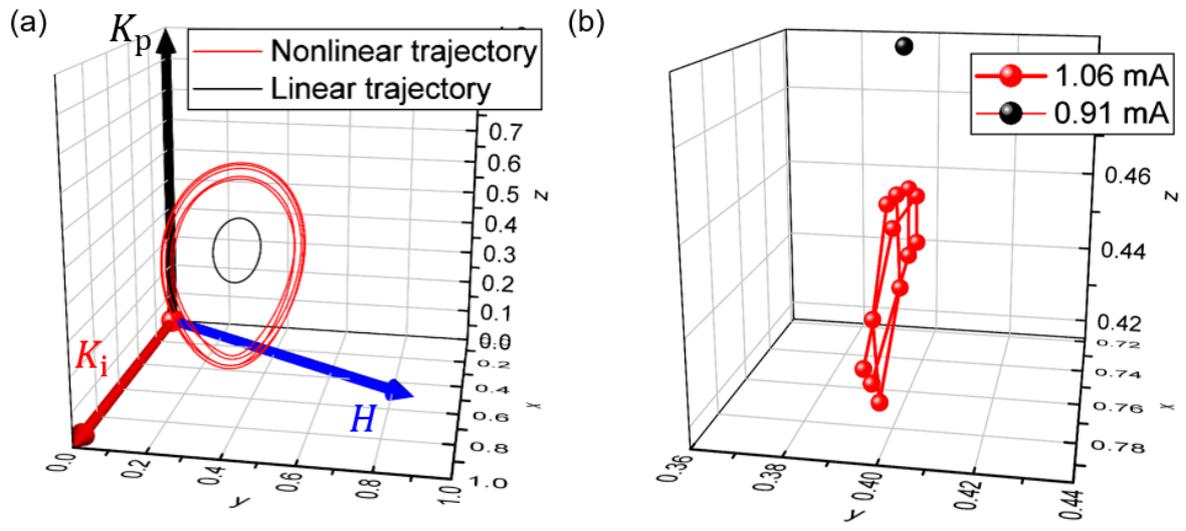

Fig. 4 Sugimoto *et al.*,

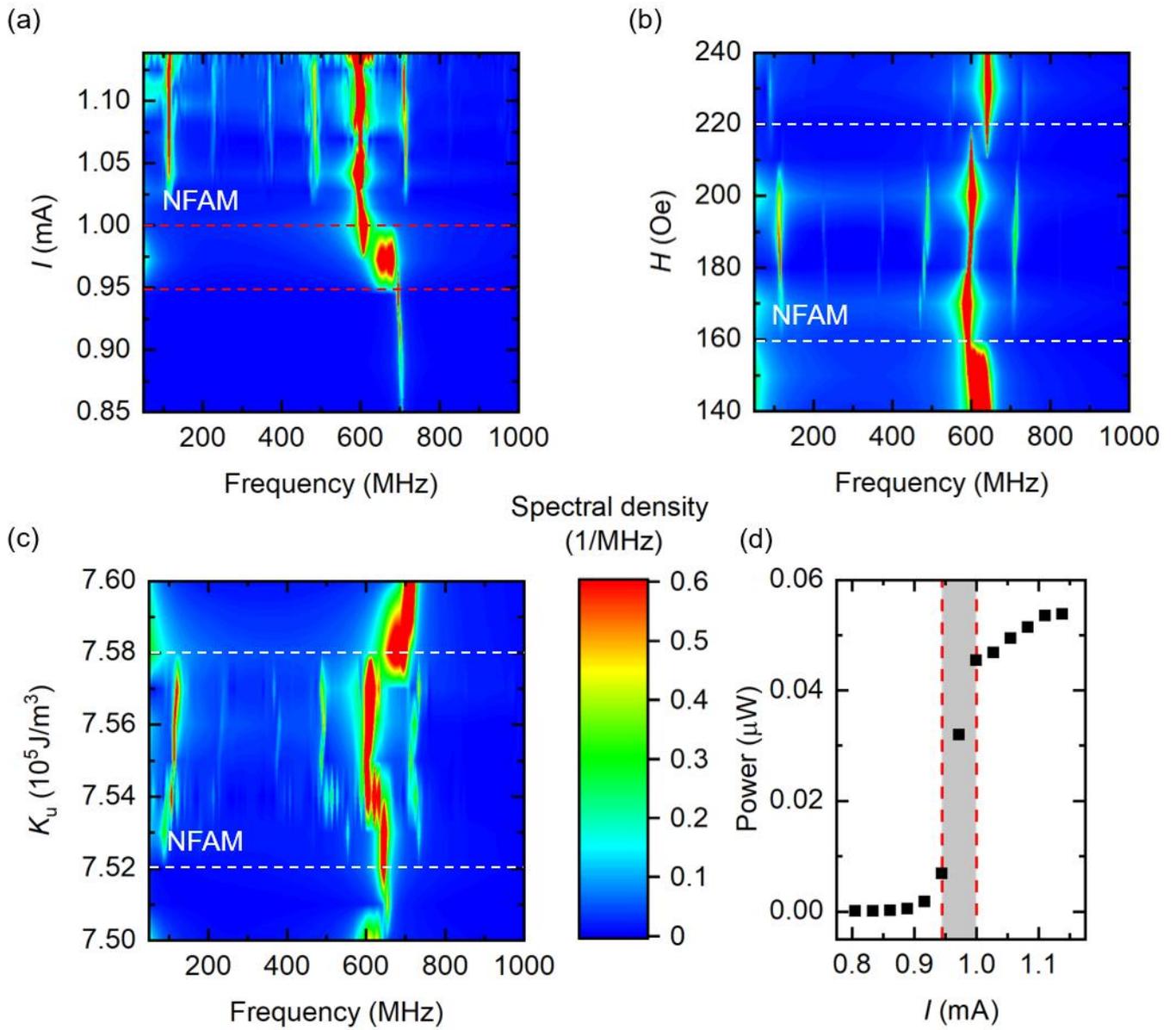

Fig. 5 Sugimoto *et al*.,